\documentclass[twoside,slac_one]{revtex4}
\usepackage{graphicx}
\usepackage{fancyhdr}
\usepackage{amsmath} 
\usepackage{bm}
\usepackage{amsxtra}
\usepackage{amssymb}
\usepackage{amsthm}
\usepackage{latexsym}
\usepackage{lscape}

\begin{document}

\title{Confronting Recent Neutrino Oscillation Data with Sterile Neutrinos}

\author{G. Karagiorgi}
\affiliation{Department of Physics, Columbia University, New York, NY, USA}

\begin{abstract}
Recent neutrino oscillation results have evoked renewed interest in sterile neutrino oscillation models. This paper reviews the data from MiniBooNE and short-baseline reactor antineutrino experiments within the context of sterile neutrinos. The results are incorporated into combined fits to test the viability of sterile neutrino oscillation models, which are later expanded to address matter effects. Finally, future experiments that can resolve the questions that have been raised are discussed.
\end{abstract}

\maketitle

\thispagestyle{fancy}

\section{\label{sec1}Sterile Neutrino Oscillation Formalism}
Sterile neutrinos refer to additional neutrino ``flavor'' (and mass) eigenstates which do not couple to the $W^\pm$ and $Z$ bosons. The additional mass eigenstates are assumed to include small admixtures of the three weakly interacting states ($\nu_e$, $\nu_{\mu}$, and $\nu_{\tau}$), so that they can be produced through mixing with the standard model neutrinos, and can therefore affect neutrino oscillations. 

In the case where only one such extra state is introduced to the standard, three-neutrino picture, the oscillation effects manifest as either $\nu_{\alpha}$ disappearance or $\nu_{\alpha}\rightarrow\nu_{\beta}$ appearance, where $\alpha,\beta=e,\mu$, or $\tau$. When the extra mass eigenstate has mass $m_4\gg m_3, m_2, m_1$, the corresponding oscillation probabilities can be approximated by the two-neutrino oscillation probabilities, 
\begin{equation}
\label{eq:aa1}
P(\stackrel{\small{(-)}}{\nu_{\alpha}}\rightarrow\stackrel{\small{(-)}}{\nu_{\alpha}})=1-\sin^22\theta_{\alpha\alpha}\sin^2(1.27\Delta m^2_{41}L/E),
\end{equation}
and 
\begin{equation}
\label{eq:mue1}
P(\stackrel{\small{(-)}}{\nu_{\alpha}}\rightarrow\stackrel{\small{(-)}}{\nu_{\beta\ne\alpha}})=\sin^22\theta_{\alpha\beta}\sin^2(1.27\Delta m^2_{41}L/E),
\end{equation}
for the case of disappearance and appearance, respectively. In the above expressions, the probability amplitudes $\sin^22\theta_{\alpha\alpha}=4|U_{\alpha4}|^2(1-|U_{\alpha4}|^2)$ and $\sin^22\theta_{\alpha\beta}=4|U_{\alpha4}|^2|U_{\beta4}|^2$ are functions which depend on the $\alpha$ and $\beta$ flavor content of the extra mass eigenstate, $|U_{\alpha4}|^2$ and $|U_{\beta4}|^2$.

In reality, more than one such state can exist. Indeed, models with two sterile neutrinos with $m_5\simeq m_4$ have been extensively studied, specifically because of the possibility of observable CP violation effects which can arise through the 4-5 interference term which appears in the appearance\footnote{Note that the disappearance probability in this scenario is insensitive to CP violation.}
oscillation probability in this scenario \cite{Karagiorgi:2006jf}. The probability can be approximated for three-neutrino mixing as
\begin{eqnarray}
P(\stackrel{\small{(-)}}{\nu_{\alpha}}\rightarrow\stackrel{\small{(-)}}{\nu_{\beta\ne\alpha}})&=&4|U_{\alpha4}|^2|U_{\beta4}|^2\sin^2(1.27\Delta m^2_{41}L/E)\nonumber \\
&~&+4|U_{\alpha5}|^2|U_{\beta5}|^2\sin^2(1.27\Delta m^2_{51}L/E)\nonumber\\
&~&+8|U_{\alpha5}||U_{\beta5}||U_{\alpha4}||U_{\beta4}|\sin(1.27\Delta m^2_{41}L/E)\sin(1.27\Delta m^2_{51}L/E)\cos(1.27\Delta m^2_{54}L/E\mp\phi_{45}),
\end{eqnarray}
where $\phi_{45}$ is a Dirac CP-violating phase, given by 
\begin{equation}
\phi_{45}=arg(U^*_{\alpha5}U_{\beta5}U_{\alpha4}U^*_{\beta4}).
\end{equation}

\section{\label{sec2}Experimental Hints}
At present, the most significant evidence for sterile neutrino oscillations at $\Delta m^2_{41}\sim1$ eV$^2$ and $\sin^22\theta_{\mu e}\sim0.005$ comes from the LSND experiment \cite{Athanassopoulos:1996jb,Athanassopoulos:1997pv,Aguilar:2001ty}. LSND was a $\pi^+\rightarrow\mu^+$ decay-at-rest experiment which employed a liquid scintillator detector to identify $\bar{\nu}_e$ events from possible $\bar{\nu}_{\mu}$ oscillations by detecting the delayed coincidence of scintillation light emitted by a positron in the $\bar{\nu}_e p\rightarrow e^+n$ reaction and light emitted by the subsequent $n$ capture. In a search for $\bar{\nu}_{\mu}\rightarrow\bar{\nu}_e$ oscillations at $L/E\sim$~0.4-1.4 m/MeV, LSND observed an excess of $\bar{\nu}_e$ events above background prediction which was consistent with oscillations described by Eq.~\ref{eq:mue1} at $3.8\sigma$.  

MiniBooNE, a pion decay-in-flight experiment, was designed to test the LSND evidence for oscillations using a higher-energy $\nu_{\mu}$ beam, and look for $\nu_e$ appearance at the same $L/E$ as LSND. In contrast to LSND, MiniBooNE employed a higher energy neutrino beam and longer baseline, and relied on the Cherenkov signature of the $e^-$ produced in a $\nu_e n\rightarrow e^-p$ interaction to identify any $\nu_e$ resulting from oscillations of $\nu_{\mu}$ in the beam. The different neutrino energy, beam, and detector systematics, as well as the different event signatures and background at MiniBooNE, allowed for an independent test of the LSND result. Under the assumption of CPT conservation, which implies that Eq.~\ref{eq:mue1} is identical for neutrinos and antineutrinos, the MiniBooNE search yielded no excess which could be consistent with the LSND signal \cite{AguilarArevalo:2007it}. Such excess would appear most noticeably in the 475-1250~MeV energy range in reconstructed electron neutrino energy. Instead, the observed $\nu_e$ data in that energy range were found consistent with background prediction. However, a $3.0\sigma$ excess of $\nu_e$ was observed at lower energy, below 475~MeV, corresponding to a significantly different $L/E$ than that of the LSND excess. The lack of excess above 475~MeV in the MiniBooNE $\nu_{\mu}\rightarrow\nu_e$ search resulted in an exclusion of the LSND 90\% confidence level (CL) allowed parameters\footnote{Using Eq.~\ref{eq:mue1} and assuming CPT conservation.} at the 98\% CL. At the same time, the low energy excess observed by MiniBooNE was found inconsistent with the single sterile neutrino oscillation scenario \cite{AguilarArevalo:2008rc}. 

Past attempts to simultaneously interpret the MiniBooNE neutrino and LSND antineutrino excesses included fits to CP-violating sterile neutrino oscillations \cite{Maltoni:2007zf,Karagiorgi:2009nb}, which are also discussed in Sec.~\ref{sec1}. Other interpretations include heavy sterile neutrino decay \cite{Gninenko:2011xa,Gninenko:2011hb,Gninenko:2009yf} and other non-standard physics scenarios such as altered neutrino dispersion relations \cite{Hollenberg:2009ak}, extra dimensions \cite{Hollenberg:2009ws}, CPT-violation \cite{Diaz:2011ia}, and non-standard neutrino interactions \cite{Akhmedov:2011zz,Nelson:2007yq}. At the same time, the possibility of mis-estimated background has been considered and ruled out by independent measurements at MiniBooNE \cite{AguilarArevalo:2008rc}, in particular for the case of NC single-photon\footnote{Cherenkov detectors such as MiniBooNE cannot distinguish a single photon from a single electron.} events, which make up the most dominant background to the MiniBooNE $\nu_e$ appearance search at low energy. However, the possibility of a new background, such as that contributed by anomaly-mediated single-photon production \cite{Harvey:2007rd}, remains a viable interpretation for the MiniBooNE low energy excess.

More recently, MiniBooNE performed a less-sensitive search for $\bar{\nu}_{\mu}\rightarrow\bar{\nu}_e$ appearance \cite{zarko} by inverting the neutrino beam polarity, resulting in a lower-intensity $\bar{\nu}_{\mu}$ beam with similar energy as the $\nu_{\mu}$ beam. This search provided a direct test of the LSND signal interpreted as $\bar{\nu}_e$ appearance with an oscillation probability given by Eq.~\ref{eq:mue1}, independent of any CPT or CP assumptions. Primarily due to low statistics, but also because of a small, $\sim1\sigma$ excess observed in the 475-1250 MeV energy range, with an oscillation-like $L/E$ dependence, the MiniBooNE $\bar{\nu}_{\mu}\rightarrow\bar{\nu}_e$ search was found to favor oscillations over the null hypothesis at the 91\% CL. At the same time, a low energy excess was also observed, though not as significant as that of the $\nu_{\mu}\rightarrow\nu_e$ search.

The above excesses are all observed in appearance channels, and if sterile neutrino oscillations were to be their underlying source, they would suggest observable $\nu_e$ and/or $\nu_{\mu}$ disappearance. Muon neutrino disappearance at $\Delta m^2\sim 0.1-10$ eV$^2$ is constrained to be less than 10\%\footnote{Approximate 90\% CL upper limit to $\sin^22\theta_{\mu\mu}$.} by short-baseline experiments including MiniBooNE/SciBooNE \cite{Mahn:2011ea}, CDHS \cite{Dydak:1983zq}, and CCFR \cite{Stockdale:1984cg}, which look for shape and normalization distortions in their respective event distributions as a function of neutrino energy, as well as MINOS \cite{Adamson:2010wi}, K2K and atmospheric neutrino experiments \cite{Maltoni:2004ei}, which look for normalization effects in their event distributions. Unless large $\nu_e$ disappearance is invoked, the $\nu_\mu$ disappearance constraints provided by the above experiments rule out the sterile neutrino oscillation interpretation of the LSND and MiniBooNE excesses. 

Electron neutrino disappearance, on the other hand, has been thought to be constrained to less than 10\% by short-baseline reactor antineutrino experiments including Bugey \cite{Declais:1994su} and CHOOZ \cite{Apollonio:2002gd}. Recently, however, a re-analysis of past reactor antineutrino experiments, including measurements performed at ILL, Bugey, ROVNO, CHOOZ, and others \cite{Mention:2011rk}, finds an overall normalization deficit observed by all those experiments, which is consistent with light sterile neutrino oscillations described by Eq.~\ref{eq:aa1}, with $\Delta m^2\sim 1$ eV$^2$ and $\sin^22\theta\sim0.1$. The oscillation fits performed in \cite{Mention:2011rk} account for a recent re-evaluation of predicted reactor antineutrino flux spectra \cite{Mueller:2011nm}, by which the new flux predictions are found to be $\sim$3\% higher than previously calculated, so that the past $\bar{\nu}_e$ flux measurements performed by the above experiments are all now systematically 3\% lower than the new predictions. The observed $\bar{\nu}_e$ deficit corresponds to $2.14\sigma$, and is referred to as the ``reactor antineutrino anomaly.'' 

This newly-found possible hint for sterile neutrino oscillations, this time in the disappearance sector,\footnote{Note that a long-standing piece of evidence for much larger $\nu_e$ disappearance at very short baselines from the Gallex and SAGE calibration source experiments has been recently excluded by a comparison of KARMEN and LSND $\nu_e$ cross section measurements to theoretical predictions \cite{Conrad:2011ce}. Those measurements, however, do not have sufficient sensitivity to address the reactor antineutrino anomaly.} has renewed the interest in global sterile neutrino oscillation fits \cite{Giunti:2011ht,Donini:2011jh,Giunti:2011gz,Kopp:2011qd}. In Sec.~\ref{sec3}, we present global oscillation fit results performed independently,\footnote{From Ref.~\cite{Karagiorgi:2009nb}, with updates to include the updated reactor and MiniBooNE results, as noted.} which confirm the findings in \cite{Kopp:2011qd}.

\section{\label{sec3}Review of Global Sterile Neutrino Oscillation Fits}

\subsection{(3+1) Oscillations}
\begin{figure}[t]
\includegraphics[width=135mm]{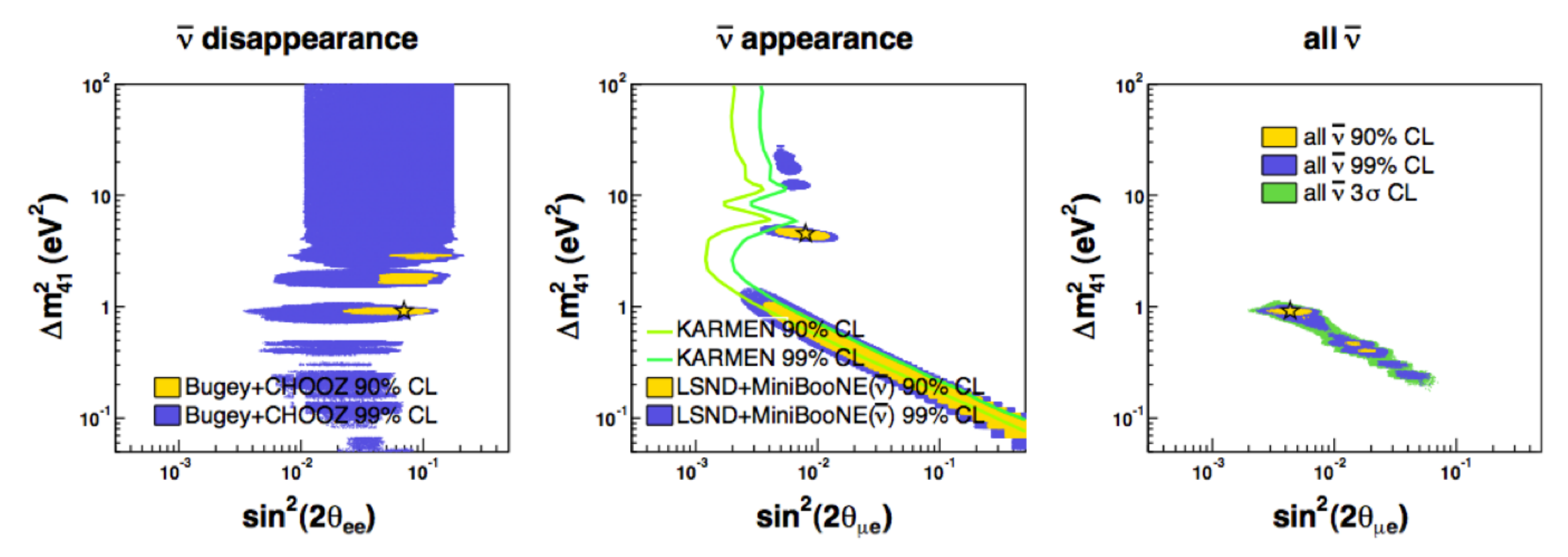}
\caption{\label{fig1}90\% and 99\% CL alowed oscillation parameter regions obtained from fits to short-baseline antineutrino-only data. Left: Fit to $\bar{\nu}_e$ disappearance data from CHOOZ and Bugey, accounting for the recent re-evaluation of predicted reactor $\bar{\nu}_e$ fluxes \cite{Mention:2011rk}. Middle: Fit to $\bar{\nu}_{\mu}\rightarrow\bar{\nu}_e$ appearance data from LSND and MiniBooNE. Also shown are the limits from the KARMEN $\bar{\nu}_{\mu}\rightarrow\bar{\nu}_e$ oscillation search. Right: Global fit to $\bar{\nu}_e$ disappearance and $\bar{\nu}_{\mu}\rightarrow\bar{\nu}_e$ appearance data sets. The global antineutrino-only fit is consistent with short-baseline (3+1) oscillations described by $\Delta m^2_{41}\sim 1$ eV$^2$ and $\sin^22\theta_{\mu}\sim0.005$ at $>3\sigma$ CL.}
\end{figure}

Under the single sterile neutrino oscillation scenario, also referred to as the (3+1) scenario, the MiniBooNE antineutrino and LSND excesses are indeed highly compatible with each other, as well as with all other short-baseline {\it antineutrino} data sets, including the KARMEN $\bar{\nu}_{\mu}\rightarrow\bar{\nu}_e$ appearance \cite{Armbruster:2002mp} and the Bugey and CHOOZ $\bar{\nu}_e$ disappearance data sets. A combined analysis of the latter is also consistent with oscillations at the $>99$\% level. A ``Parameter Goodness-of-fit'' (PG) test \cite{Maltoni:2003cu} results in a 22\% compatibility among all above antineutrino experiments. The level of compatibility among antineutrino data sets is illustrated qualitatively in Fig.~\ref{fig1}.

\begin{figure}[t]
\includegraphics[width=60mm]{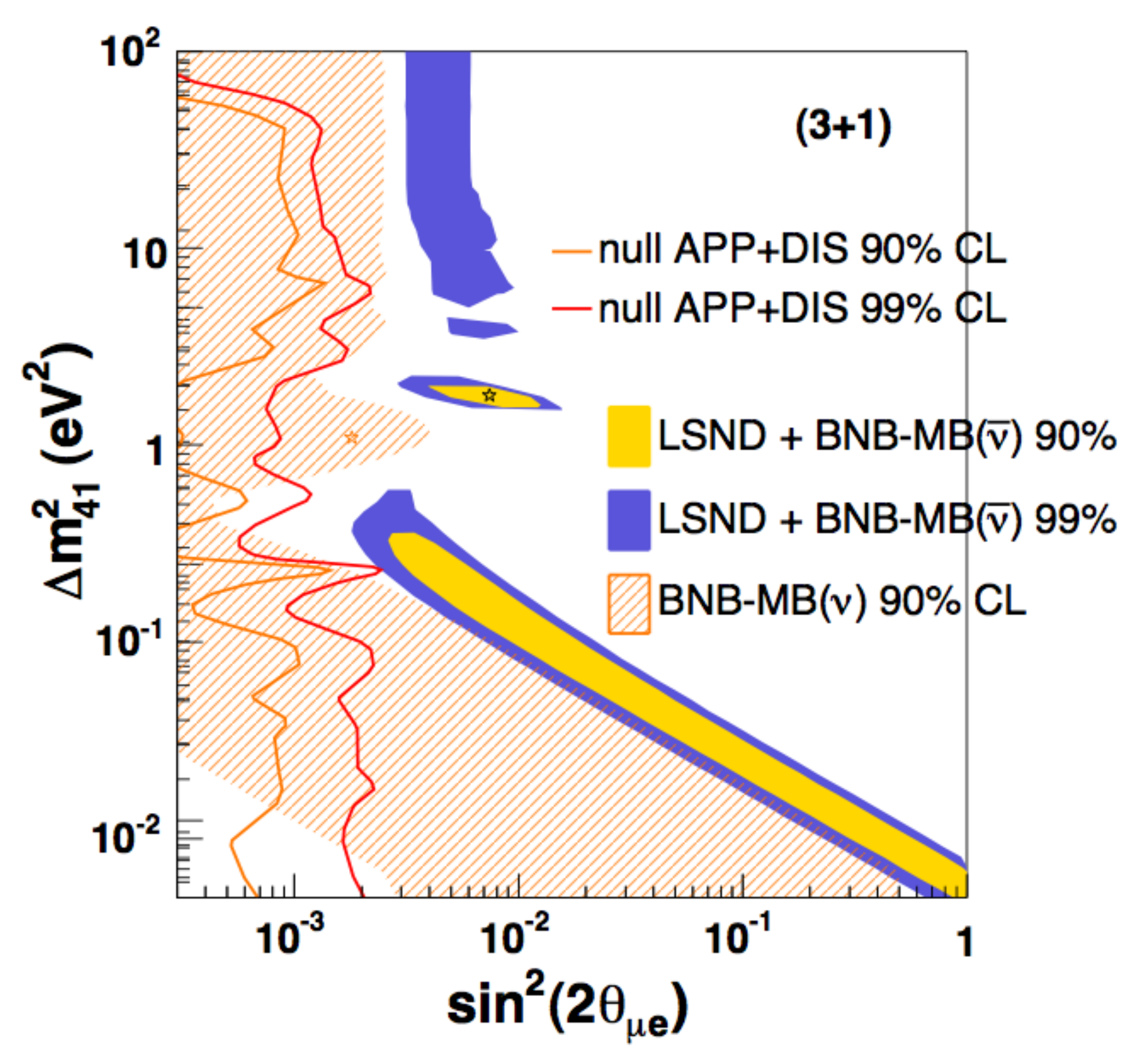}
\caption{\label{fig2}90\% and 99\% CL alowed oscillation parameters (filled regions) and limits (lines) obtained from fits to short-baseline neutrino and antineutrino data. The solid lines correspond to 90\% and 99\% CL exclusion limits obtained from combined fits to null atmospheric $\nu_{\mu}$ disappearance and null short-baseline experiments. The filled region corresponds to the jointly-allowed LSND and MiniBoo\
NE antineutrino fit oscillation parameters, while the filled hashed region corresponds to the parameter region (marginally) allowed by MiniBooNE neutrino data. The figure is based on the analysis performed in \cite{Karagiorgi:2009nb}, using an older, lower-statistics MiniBooNE antineutrino data set, and reactor $\bar{\nu}_e$ disappearance data sets prior to the recent reactor flux re-analysis. Note that the updates do not significantly alter the conclusions drawn from this picture.}
\end{figure}

Of course, except for the purpose of making an empirical observation, namely the good compatibility of all antineutrino data sets under this simple oscillation assumption, one cannot ignore the constraints from the $\nu_{\mu}$ disappearance experiments, as well as additional $\nu_e$ appearance constraints from NOMAD \cite{Astier:2003gs} and the MiniBooNE $\nu_{\mu}\rightarrow\nu_e$ search. Under a CPT conserving (3+1) scenario, neutrino and antineutrino oscillations are identical, and therefore constraints from neutrino data sets are directly applicable to the global fits considered under this scenario. The level of constraint provided by the neutrino data sets is illustrated in Fig.~\ref{fig2}. As suggested by the figure, it is impossible to reconcile all short-baseline results under a CPT-conserving (3+1) scenario. In this case, the compatibility of all short-baseline datasets returned by a PG test is 0.11\%, essentially ruling out this oscillation hypothesis.

\subsection{(3+2) Oscillations}
\begin{figure}[t]
\includegraphics[width=70mm]{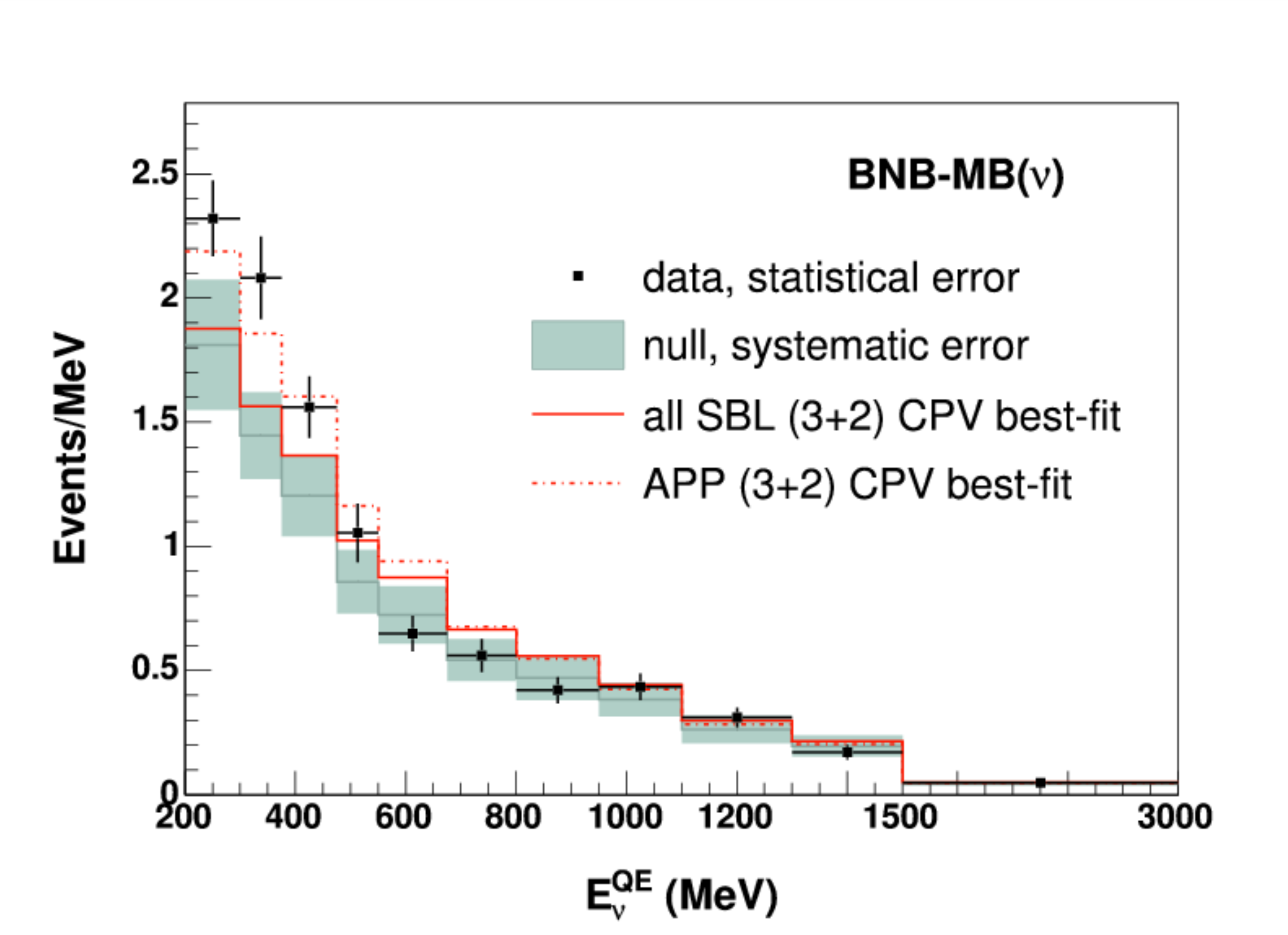}\includegraphics[width=70mm]{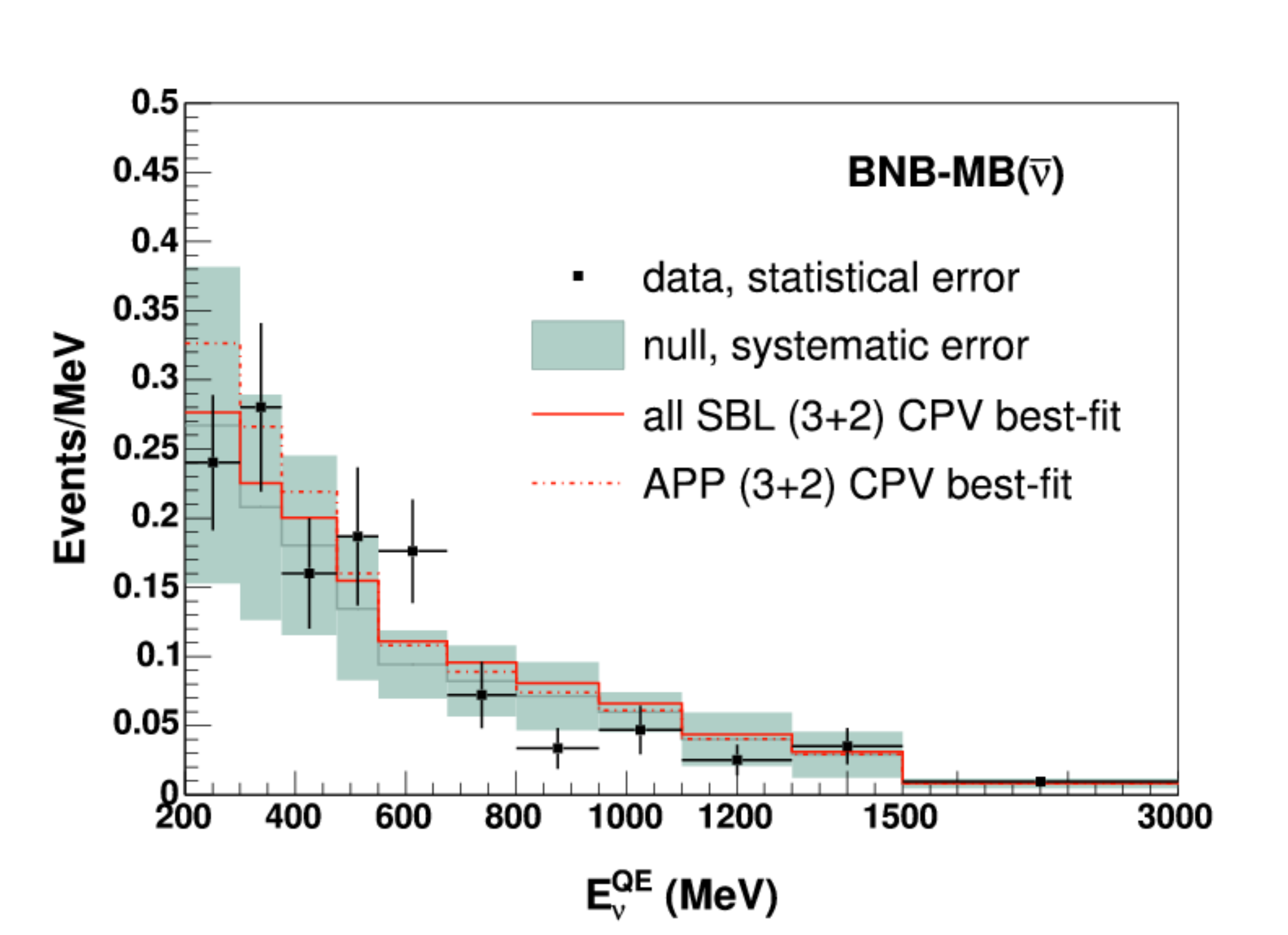}
\caption{\label{fig3}The MiniBooNE $\nu_e$ (left) and $\bar{\nu}_e$ (right) event distributions versus reconstructed neutrino energy, from \cite{Karagiorgi:2009nb}. The null (no-oscillations) predictions are shown in green, with gray systematic error bands. The observed data are shown in black points, with statistical error bars. The signal predictions from a joint (3+2) CP-violating fit to all appearance (dashed red) and all appearance and disappearance (solid red) data sets are overlaid. The (3+2) CP-violating fit to all data sets fails to explain the observed excess signatures at MiniBooNE.}
\end{figure}

As mentioned earlier, a (3+2) oscillation scenario is attractive because it offers the possibility of using CP violation to account for the observed differences in---what could be evidence of---short-baseline neutrino and antineutrino oscillations. Indeed, a combined analysis of all {\it appearance} data sets under the (3+2) scenario yields high compatibility \cite{Karagiorgi:2009nb} if one allows for large CP violation; furthermore, it predicts an excess of low energy events at MiniBooNE in both neutrino and antineutrino running, which can account for the observed excesses, as illustrated in Fig.~\ref{fig3}. However, when disappearance data sets all included in the fits, the compatibility reduces dramatically, and one finds large tension among neutrino and antineutrino data sets.

\section{Expanding the Formalism to Account for Matter-like Effects}
\label{sec4}
The present (3+1) and (3+2) sterile neutrino oscillation fits reveal that neutrino and antineutrino oscillation signatures which may be responsible for the LSND, MiniBooNE, and reactor antineutrino anomalies are incompatible with other, primarily neutrino, null results. The neutrino and antineutrino data sets show differences which cannot be explained away by just CP violation. That observation, as well as a (now essentially resolved) hint from the MINOS $\bar{\nu}_{\mu}$ disappearance search suggesting CPT-violating differences in neutrino and antineutrino atmospheric oscillations \cite{Danko:2009qw,minosnubar}, has prompted the theoretical community to consider even less standard scenarios, beyond the ``reference picture'' of CPT-conserving 3+N oscillations \cite{everett}. An example of such scenarios, involving sterile neutrinos and non-standard matter-like effects \cite{inprep}, is discussed below.

\subsection{Extended Oscillation Formalism}
The hypothesis being investigated in this section is one where the appearance signals from MiniBooNE neutrino and antineutrino running and LSND are due to a (3+1) oscillation scenario, where $\stackrel{\small{(-)}}{\nu_s}$ experience a matter-like potential $V_s=\pm A_s$. The resulting appearance oscillation probability under this scenario is given by
\begin{equation}
\label{eq:mfx}
P(\stackrel{\small{(-)}}{\nu_{\mu}}\rightarrow\stackrel{\small{(-)}}{\nu_e})=\sin^22\theta_M\sin^2(1.27\Delta m^2_ML/E),
\end{equation}
where $\sin^22\theta_M$ and $\Delta m^2_M$ are the effective mixing parameters due to the presence of $V_s\ne0$. The oscillation probability has a form identical to that of a (3+1) scenario, except that the {\it effective} oscillation parameters are dependent both on the underlying ``vacuum'' (3+1) oscillation parameters and the neutrino energy and $V_s$ potential, according to
\begin{equation}
\Delta m^2_M=\Delta m^2_{41}+2EV_s,
\end{equation}
and
\begin{eqnarray}
\sin^2\theta_M&=&\frac{16(\Delta m^2_{41})^4|U_{e4}|^2|U_{\mu4}|^2|U_{s4}|^4}{((\Delta m^2_{41}-2EV_s)^2+8EV_s\Delta m^2_{41}|U_{s4}|^2)}\cdot\nonumber\\
&~&\frac{1}{(2EV_s-\Delta m^2_{41}(1-2|U_{s4}|^2)+\sqrt{(\Delta m^2_{41}-2EV_s)^2+8EV_s\Delta m^2_{41}|U_{s4}|^2})^2}.
\end{eqnarray}
The dependence on $EV_s$, where $V_s=\pm A_s$, suggests different and also energy-dependent effective oscillation parameters for neutrinos versus antineutrinos. More specifically, for a set of vacuum oscillation parameters $\Delta m^2_{41}$, $|U_{\mu4}|^2$, and $|U_{e4}|^2$, a non-zero $A_s$ will lead to opposing deviations from the vacuum oscillation parameters for neutrinos versus antineutrinos, as well as resonance effects which shift to lower neutrino and antineutrino energies as $A_s$ increases.

The following subsection presents results from a joint fit to this scenario performed using the MiniBooNE and LSND appearance data sets, assuming $\nu_{\mu}\rightarrow\nu_e$ and $\bar{\nu}_{\mu}\rightarrow\bar{\nu}_e$ oscillations described by Eq.~\ref{eq:mfx}. During the fit, the vacuum oscillation parameters, $\Delta m^2_{41}$, $|U_{e,\mu4}|^2$, and $A_s$, are varied within 0.01-100~eV$^2$, 0-0.05, and $10^{-13}$-$10^{-9}$~eV, respectively.\footnote{The choice of $A_s$ range is driven by short-baseline data.}

\subsection{Fit Results}
\begin{figure}
\includegraphics[width=100mm]{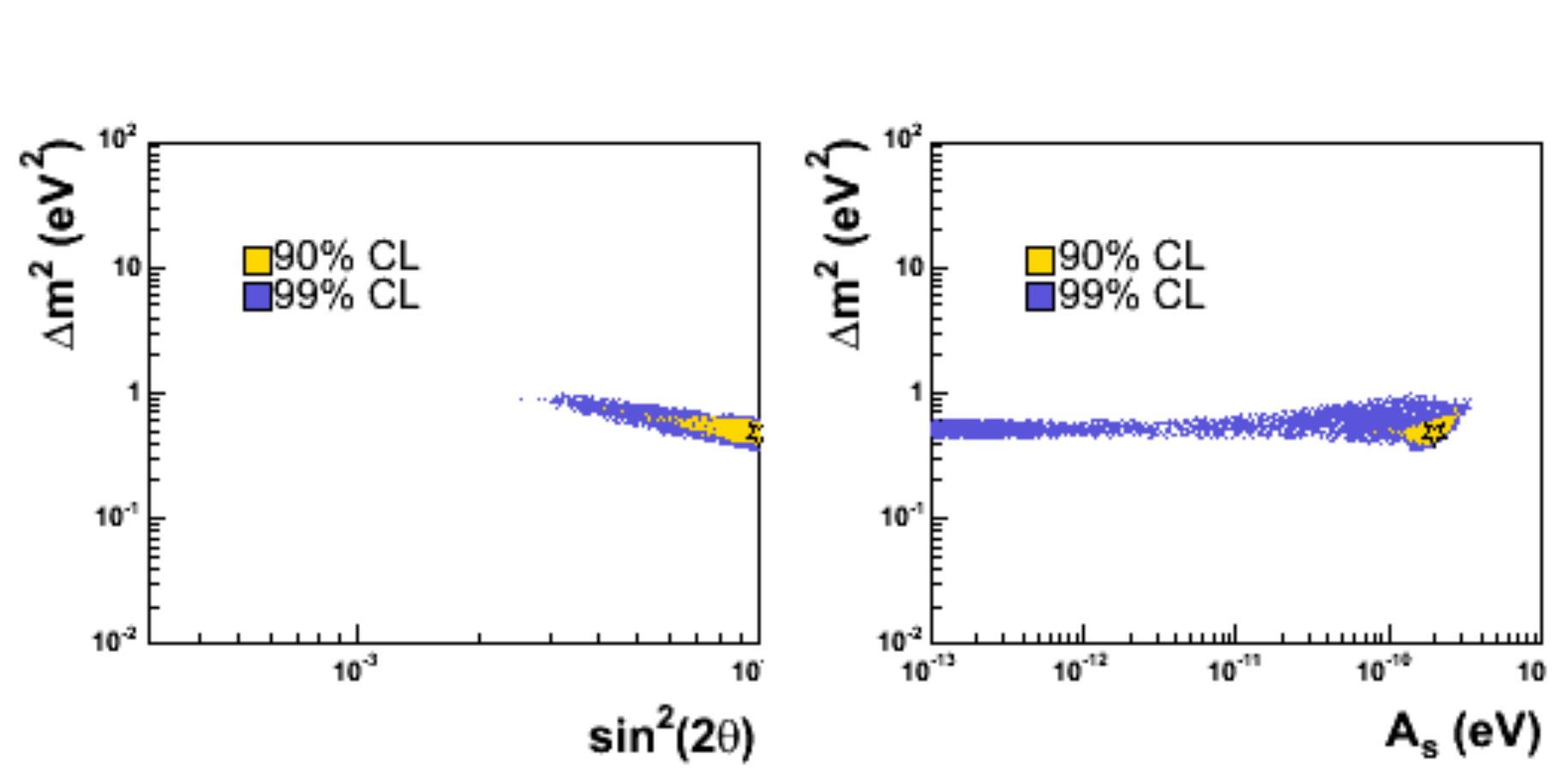}
\caption{\label{fig4}The 90\% and 99\% CL allowed vacuum oscillation parameters $\Delta m^2_{41}$ and $\sin^22\theta_{\mu e}=4|U_{e4}|^2|U_{\mu4}|^2$ and $A_s$ potential obtained from a joint (3+1) with matter-like effects fit to the MiniBooNE and LSND data sets. The star indicates the best-fit parameters.}
\end{figure}

Relative to a (3+1) fit, the compatibility among MiniBooNE and LSND data sets when one allows for non-zero $A_s$ increases from 2.3\% to 17.4\%. The allowed oscillation parameters are shown in Fig.~\ref{fig4}. The returned best-fit vaccum oscillation parameters corespond to $\Delta m^2_{41}=0.47$ eV$^2$ and $\sin^22\theta_{\mu e} =4|U_{e4}|^2|U_{\mu4}|^2=0.01$, and the best-fit $A_s$ value corresponds to 2$\times$10$^{-10}$ eV. Note that the best-fit $A_s$ value is significantly larger than $A_e=\sqrt{2}G_Fn_e\sim10^{-13}$ eV for standard matter effects.

\begin{figure}
\includegraphics[width=70mm]{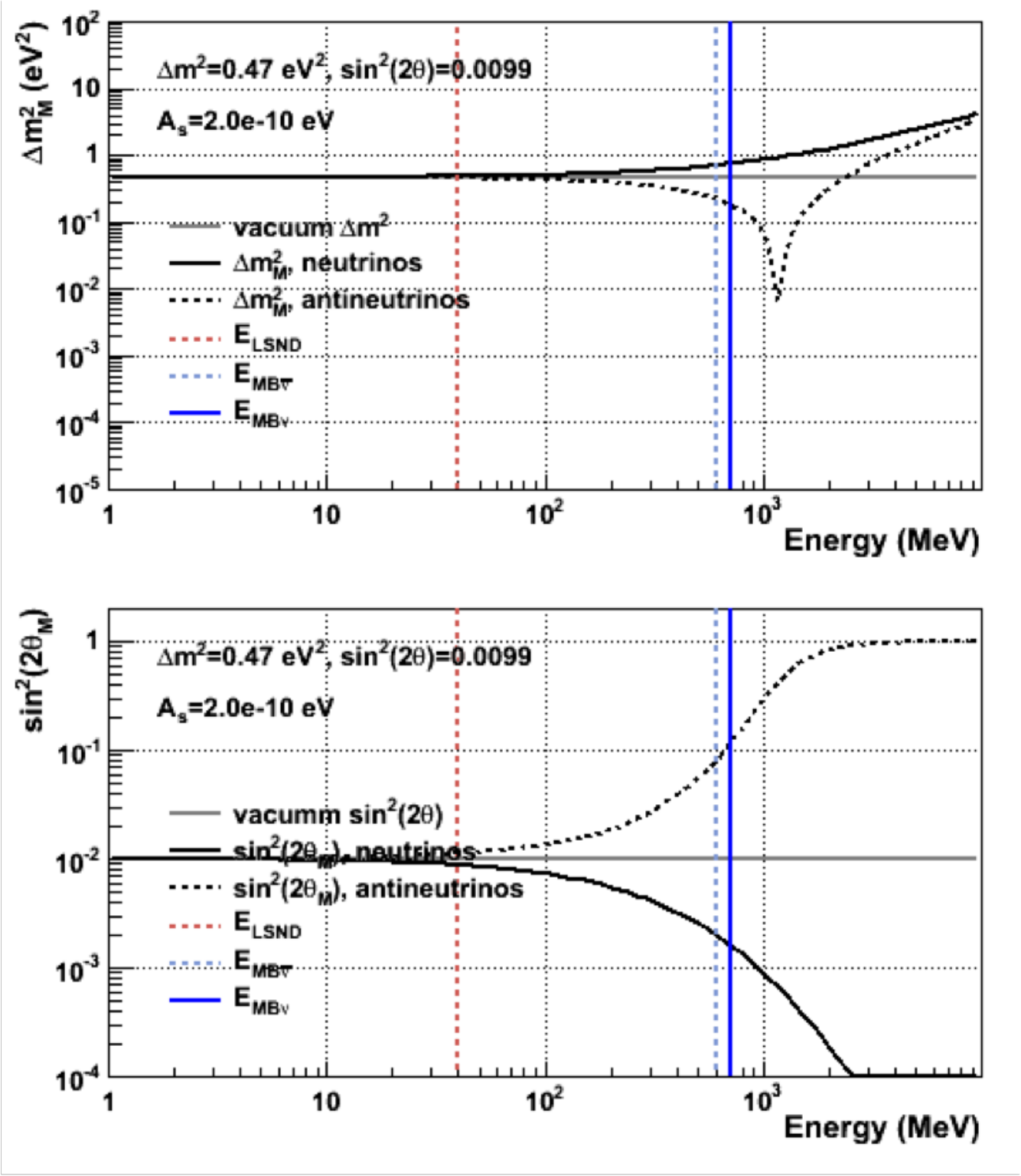}
\caption{\label{fig5}Energy dependence of the effective mixing parameters corresponding to the best-fit parameters in Fig.~\ref{fig5}. The gray flat lines correspond to the best-fit vacuum oscillation parameters, while the solid (dashed) black lines correspond to the effective best-fit oscillation parameters for neutrinos (antineutrinos). The vertical lines approximate the mean neutrino energy in the MiniBooNE and LSND antineutrino (dashed) and neutrino (solid) appearance searches. At low energies, the effective oscillation parameters reduce to the vacuum ones, and no differences in neutrino versus antineutrino oscillations are expected; at higher energies, one sees increasing deviations from the vacuum case as well as increasingly larger differences between neutrinos and antineutrinos. The plot suggests observable effects at high-energy ($>1GeV$) antineutrino data sets.}
\end{figure}

The effective mixing parameters corresponding to the best-fit vacuum oscillation parameters and $A_s$ are shown as a function of neutrino (and antineutrino) energy in Fig.~\ref{fig5}. The effective parameters at lower energies, below 10 MeV, reduce to the vacuum (3+1) oscillation parameters, and could likely accommodate the reactor antineutrino anomaly (3+1) oscillation interpretation. At higher energy, however, the effective mixing parameters suggest observable effects in the MINOS, K2K, IceCube, and atmospheric antineutrino measurements. Those effects are currently being investigated \cite{inprep}.

\section{\label{sec5}Future Tests}
Experiments which aim to address the short-baseline anomalies discussed in Sec.~\ref{sec2} include both appearance and disappearance experiments. Newly proposed disappearance experiments include reactor antineutrino disappearance searches at very short baselines of the order of 10 m or less, where the $L/E$ dependence of possible oscillation effects can be directly observed rather than inferred due to an overall normalization deficit, as is presently the experimental case. Such experiments include NUCIFER \cite{Porta:2010zz} and SCRAAM \cite{llnl}. Similar searches are proposed for very short baseline $\nu_e$ and $\bar{\nu}_e$ disappearance using MegaCurie-strength radioactive sources placed either inside or in close proximity to a neutrino detector, such that oscillation ``wiggles'' can be fully observed within the detector spatial dimensions. An example of such experiment is discussed in \cite{Formaggio:2011jt}.

Of course, appearance-style short-baseline oscillation experiments are also being proposed, where particular effort is being made to focus on experiments with the highest possible (ideally, 5$\sigma$) sensitivity to $\bar{\nu}_{\mu}\rightarrow\bar{\nu}_e$ and/or $\nu_{\mu}\rightarrow\nu_e$ appearance. Those proposals include short-baseline oscillation searches at OscSNS, BooNE, Gd-doped Super-K running with a decay-at-rest neutrino source, and coherent scattering experiments with a decay-at-rest neutrino source \cite{Ray:2008ea,Stancu:2009vq,Agarwalla:2010zu,figueroa}.

At the same time, several presently running and planned experiments aim to shed more light onto the picture. Specifically, MiniBooNE antineutrino running has been ongoing, aiming to nearly double the present antineutrino statistics to a sample corresponding to 1.5$\times$10$^{20}$ POT by mid 2012. The higher-statistics search would still be statistics- rather than systematics-limited, but could nevertheless either strengthen the significance of the present antineutrino result to $\sim3\sigma$, or reduce it to the $<1\sigma$ level. 

A more definitive test can be performed by MicroBooNE \cite{Chen:2007zz}, which is an experiment under construction which will run in the same neutrino beamline as MiniBooNE, at roughly the same baseline. In contrast to MiniBooNE, MicroBooNE uses a 60-ton (fiducial mass) liquid argon time projection chamber (LArTPC) detector, which has a significantly higher detection and PID efficiency than MiniBooNE. MicroBooNE presently plans to run in a neutrino beam, so that it can perform a definitive test of the MiniBooNE low energy excess. Specifically, MicroBooNE's single-electron versus single-photon high differentiation capability grants a $>5\sigma$ sensitivity to a single-electron interpretation of the low energy excess events, and a $>4\sigma$ sensitivity to a single-photon interpretation. MicroBooNE is also in position to perform a $\nu_{\mu}\rightarrow\nu_e$ appearance search; however, its relatively small size limits the statistical sensitivity of such search to a level which would only be comparable to the MiniBooNE sensitivity. A proposed future MicroBooNE run with a second, larger LArTPC detector \cite{roxanne}, however, could provide a $>5\sigma$ sensitivity in both neutrino running, as well as antineutrino running. A similar oscillation search involving two LArTPC detectors at the CERN-PS beam has also been proposed by an independent group \cite{Baibussinov:2009tx}.
 
\section{\label{sec6}Summary}
Over the last five years, a number of hints have surfaced which are in support of the LSND evidence for sterile neutrino oscillations. Those hints come from $\nu_{\mu}\rightarrow\nu_e$ and $\bar{\nu}_{\mu}\rightarrow\bar{\nu}_e$ appearance measurements at MiniBooNE, and a combined analysis of past $\bar{\nu}_e$ disappearance measurements performed at reactors at short baselines. At the same time, however, constraints from $\nu_{\mu}$ disappearance measurements by MINOS, K2K, MiniBooNE, and other experiments which show no evidence for such oscillations, place stringent limits to those models. Over the past five years, global fit results have revealed increasing tension between neutrino and antineutrino data sets when tested under the sterile neutrino oscillation hypothesis. The apparent differences are such that they cannot be accommodated by any CP violation allowed within the model. Non-standard interactions and matter-like effects, such as of the type discussed in Sec.~\ref{sec5}, generally induce effects which look like CPT violation, and therefore show promise in being able to simultaneously explain the LSND, MiniBooNE, and reactor anomalies. Those effects have received increased interest within the community over the past year, and should be further developed and tested. An example of such model has been presented in this paper. While the model yields a satisfactory fit to the data, it requires a significantly large ``matter-like'' potential $V_s=\pm A_s$, experienced only by sterile neutrinos, which invites theoretical interpretation. At the same time, it suggests observable effects in higher-energy neutrino and antineutrino experiments. Therefore, the data provided by MINOS, IceCube and T2K will be crucial in addressing the viability of the model. On the other hand, high-precision, very short baseline experiments at reactors should be able to directly test the sterile neutrino oscillation interpretation for the reactor anomaly. Several proposals for new appearance-style experiments are also being developed, with the goal of a definitive, $>5\sigma$ test of the LSND and MiniBooNE results.

\begin{acknowledgments}
The author acknowledges J.~M.~Conrad and M.~H.~Shaevitz for their collaboration on the phenomenological study presented in Sec.~\ref{sec5}.
\end{acknowledgments}

\bigskip 


\begin{thebibliography}{99}   

\bibitem{Karagiorgi:2006jf}
  G.~Karagiorgi, A.~Aguilar-Arevalo, J.~M.~Conrad, M.~H.~Shaevitz, K.~Whisnant, M.~Sorel, V.~Barger,
  Phys.\ Rev.\  {\bf D75}, 013011 (2007).

\bibitem{Athanassopoulos:1996jb}
  C.~Athanassopoulos {\it et al.} [ LSND Collaboration ],
  Phys.\ Rev.\ Lett.\  {\bf 77}, 3082-3085 (1996).

\bibitem{Athanassopoulos:1997pv}
  C.~Athanassopoulos {\it et al.} [ LSND Collaboration ],
  Phys.\ Rev.\ Lett.\  {\bf 81}, 1774-1777 (1998).

\bibitem{Aguilar:2001ty}
  A.~Aguilar {\it et al.} [ LSND Collaboration ],
  Phys.\ Rev.\  {\bf D64}, 112007 (2001).

\bibitem{AguilarArevalo:2007it}
  A.~A.~Aguilar-Arevalo {\it et al.} [ The MiniBooNE Collaboration ],
  Phys.\ Rev.\ Lett.\  {\bf 98}, 231801 (2007).

\bibitem{AguilarArevalo:2008rc}
  A.~A.~Aguilar-Arevalo {\it et al.} [ MiniBooNE Collaboration ],
  Phys.\ Rev.\ Lett.\  {\bf 102}, 101802 (2009).

\bibitem{Maltoni:2007zf}
  M.~Maltoni, T.~Schwetz,
  Phys.\ Rev.\  {\bf D76}, 093005 (2007).
  [arXiv:0705.0107 [hep-ph]].

\bibitem{Karagiorgi:2009nb}
  G.~Karagiorgi, Z.~Djurcic, J.~M.~Conrad, M.~H.~Shaevitz, M.~Sorel,
  Phys.\ Rev.\  {\bf D80}, 073001 (2009).

\bibitem{Gninenko:2011xa}
  S.~N.~Gninenko,
  Phys.\ Rev.\  {\bf D83}, 093010 (2011).

\bibitem{Gninenko:2011hb}
  S.~N.~Gninenko,
  [arXiv:1107.0279 [hep-ph]].

\bibitem{Gninenko:2009yf}
  S.~N.~Gninenko, D.~S.~Gorbunov,
  Phys.\ Rev.\  {\bf D81}, 075013 (2010).

\bibitem{Hollenberg:2009ak}
  S.~Hollenberg, O.~Micu, H.~Pas,
  Prog.\ Part.\ Nucl.\ Phys.\  {\bf 64}, 193-195 (2010).

\bibitem{Hollenberg:2009ws}
  S.~Hollenberg, O.~Micu, H.~Pas, T.~J.~Weiler,
  Phys.\ Rev.\  {\bf D80}, 093005 (2009).

\bibitem{Diaz:2011ia}
  J.~S.~Diaz, A.~Kostelecky,
  [arXiv:1108.1799 [hep-ph]].

\bibitem{Akhmedov:2011zz}
  E.~K.~.Akhmedov, T.~Schwetz,
  Nucl.\ Phys.\ Proc.\ Suppl.\  {\bf 217}, 217-219 (2011).

\bibitem{Nelson:2007yq}
  A.~E.~Nelson, J.~Walsh,
  Phys.\ Rev.\  {\bf D77}, 033001 (2008).

\bibitem{Harvey:2007rd}
  J.~A.~Harvey, C.~T.~Hill, R.~J.~Hill,
  Phys.\ Rev.\ Lett.\  {\bf 99}, 261601 (2007).

\bibitem{zarko}
  Z.~Pavlovic, {\it these proceedings}.

\bibitem{Mahn:2011ea}
  K.~B.~M.~Mahn {\it et al.} [ SciBooNE and MiniBooNE Collaboration ],
  [arXiv:1106.5685 [hep-ex]].

\bibitem{Dydak:1983zq}
  F.~Dydak, G.~J.~Feldman, C.~Guyot, J.~P.~Merlo, H.~J.~Meyer, J.~Rothberg, J.~Steinberger, H.~Taureg {\it et al.},
  Phys.\ Lett.\  {\bf B134}, 281 (1984).
  
\bibitem{Stockdale:1984cg}
  I.~E.~Stockdale, A.~Bodek, F.~Borcherding, N.~Giokaris, K.~Lang, D.~Garfinkle, F.~S.~Merritt, M.~Oreglia {\it et al.},
  Phys.\ Rev.\ Lett.\  {\bf 52}, 1384 (1984).

\bibitem{Adamson:2010wi}
  P.~Adamson {\it et al.} [ The MINOS Collaboration ],
  Phys.\ Rev.\  {\bf D81}, 052004 (2010).
  
\bibitem{Maltoni:2004ei}
  M.~Maltoni, T.~Schwetz, M.~A.~Tortola and J.~W.~F.~Valle,
  New J.\ Phys.\  {\bf 6}, 122 (2004).

\bibitem{Declais:1994su}
  Y.~Declais, J.~Favier, A.~Metref, H.~Pessard, B.~Achkar, M.~Avenier, G.~Bagieu, R.~Brissot {\it et al.},
  Nucl.\ Phys.\  {\bf B434}, 503-534 (1995).

\bibitem{Apollonio:2002gd}
  M.~Apollonio {\it et al.} [ CHOOZ Collaboration ],
  Eur.\ Phys.\ J.\  {\bf C27}, 331-374 (2003).

\bibitem{Mention:2011rk}
  G.~Mention, M.~Fechner, T.~.Lasserre, T.~.A.~Mueller, D.~Lhuillier, M.~Cribier, A.~Letourneau,
  Phys.\ Rev.\  {\bf D83}, 073006 (2011).

\bibitem{Mueller:2011nm}
  T.~.A.~Mueller, D.~Lhuillier, M.~Fallot, A.~Letourneau, S.~Cormon, M.~Fechner, L.~Giot, T.~Lasserre {\it et al.},
  Phys.\ Rev.\  {\bf C83}, 054615 (2011).

\bibitem{Conrad:2011ce}
  J.~M.~Conrad, M.~H.~Shaevitz,
  [arXiv:1106.5552 [hep-ex]].

\bibitem{Giunti:2011ht}
  C.~Giunti,
  [arXiv:1106.4479 [hep-ph]].

\bibitem{Donini:2011jh}
  A.~Donini, P.~Hernandez, J.~Lopez-Pavon, M.~Maltoni,
  JHEP {\bf 1107}, 105 (2011).

\bibitem{Giunti:2011gz}
  C.~Giunti, M.~Laveder,
  [arXiv:1107.1452 [hep-ph]].

\bibitem{Kopp:2011qd}                   
  J.~Kopp, M.~Maltoni, T.~Schwetz,
  Phys.\ Rev.\ Lett.\  {\bf 107}, 091801 (2011).

\bibitem{Armbruster:2002mp}
  B.~Armbruster {\it et al.} [ KARMEN Collaboration ],
  Phys.\ Rev.\  {\bf D65}, 112001 (2002).

\bibitem{Maltoni:2003cu}
  M.~Maltoni, T.~Schwetz,
  Phys.\ Rev.\  {\bf D68}, 033020 (2003).

\bibitem{Astier:2003gs}
  P.~Astier {\it et al.} [ NOMAD Collaboration ],
  Phys.\ Lett.\  {\bf B570}, 19-31 (2003).

\bibitem{Danko:2009qw}
  I.~Danko [ MINOS Collaboration ],
  [arXiv:0910.3439 [hep-ex]].

\bibitem{minosnubar}
X.~Huang, {\it these proceedings}.

\bibitem{everett}
L.~Everett, {\it these proceedings}.

\bibitem{inprep}
G.~Karagiorgi, J.~M.~Conrad, M.~H.~Shaevitz, {\it in preparation}.

\bibitem{Porta:2010zz}
  A.~Porta [ Nucifer Collaboration ],
  J.\ Phys.\ Conf.\ Ser.\  {\bf 203}, 012092 (2010).
  
\bibitem{llnl}
  N.~Bowden, {\it Sterile Neutrinos at the Crossroads Workshop, 2011}.

\bibitem{Formaggio:2011jt}
  J.~A.~Formaggio, E.~Figueroa-Feliciano, A.~J.~Anderson,
  [arXiv:1107.3512 [hep-ph]].

\bibitem{Ray:2008ea}
  H.~Ray  [OscSNS Collaboration],
  J.\ Phys.\ Conf.\ Ser.\  {\bf 136}, 022029 (2008).

\bibitem{Stancu:2009vq}
  I.~Stancu {\it et al.},
  arXiv:0910.2698 [hep-ex].

\bibitem{Agarwalla:2010zu}
  S.~K.~Agarwalla, P.~Huber,
  Phys.\ Lett.\  {\bf B696}, 359-361 (2011).

\bibitem{figueroa}
  E.~Figueroa-Feliciano, {\it these proceedings.}

\bibitem{Chen:2007zz}
  H.~Chen {\it et al.} [ MicroBooNE Collaboration ], MicroBooNE Proposal.

\bibitem{roxanne}
  R.~Guenette, {\it 2nd International Workshop towards the Giant Liquid Argon Charge Imaging Experiment (GLA2011)}.

\bibitem{Baibussinov:2009tx}
  B.~Baibussinov, E.~Calligarich, S.~Centro, D.~Gibin, A.~Guglielmi, F.~Pietropaolo, C.~Rubbia, P.~Sala,
  [arXiv:0909.0355 [hep-ex]].

\end{thebibliography}
\end{document}